\documentstyle[psfig,amssymb]{mn}
\title[The XMM-Newton/2dF survey I: X-ray properties of normal galaxies]
{The XMM-Newton/2dF survey I: X-ray properties of normal galaxies}

\author[A. Georgakakis et al.] {A. Georgakakis$^{1}$\thanks{email: age@astro.noa.gr}, 
  I. Georgantopoulos$^{1}$,  G. C. Stewart$^2$,
  T. Shanks$^3$, B.J. Boyle$^4$ \\ \\ 
  $^1$ Institute of Astronomy \& Astrophysics, National Observatory of
  Athens, I. Metaxa \& B. Pavlou, Penteli, 15236, Athens, Greece \\ 
  $^2$ Department of Physics and Astronomy, University of Leicester
  Leicester LE1 7RH \\
  $^3$ Physics Department, University of Durham,  Science Labs,
  South Road, Durham, DH1 3LE \\
  $^4$ Anglo-Australian Observatory, PO Box 296, Epping, NSW 2121, Australia \\
}
\begin{document}
\maketitle  

\begin{abstract}
This paper explores the X-ray properties of `normal' galaxies using a
shallow (2-10\,ksec) XMM-{\it Newton} survey covering an area 
of $\approx 1.5\, \rm deg^{2}$. The X-ray survey overlaps with the 2dF
Galaxy  Redshift Survey. Compared with previous studies this has the
advantage that high quality spectra and spectral classifications
(early, late type) exist for all galaxies to $b_j=19.4$. Moreover,
sources with  optical spectra revealing powerful AGNs can easily be
discarded from the normal galaxy sample used here.  In particular, we
present stacking analysis results for  about 200 galaxies from the  2dF
Galaxy  Redshift Survey at a mean redshift of $z\approx0.1$.   
We detect a strong signal for the whole sample ($\approx6\,\sigma$) in
the soft  0.5-2\,keV band corresponding to a flux of $\approx 7 
\times10^{-16}\, \rm erg \, sec^{-1} \, cm^{-2}$ and a luminosity of
$\approx 2\times 10^{40}\rm \,  erg \, sec^{-1}$. A statistically
significant  signal is
also detected for both the early and late galaxy sub-samples
with X-ray luminosities of $\approx 3 \times 10^{40}$ and $\approx
5 \times  10^{39}\, \rm erg \, sec^{-1}$ respectively. In contrast, no
signal is detected in the hard 2-8\,keV band for any of the above    
samples. The mean $L_X/L_B$ ratio of the spiral galaxy sample is found
to be consistent with both local ($<100$\,Mpc) and distant
($z\approx1$) samples (after accounting for differences in $L_B$)
suggesting little or no evolution of the X-ray emission 
mechanisms relative to the optical. 
The 0.5-2\,keV XRB contribution of the spiral galaxy sub-sample at
$z\approx0.1$  is estimated to be 0.4 per cent in broad agreement with the
XRB fractions estimated in previous studies. 
Assuming that star-forming galaxies evolve with
redshift as $(1+z)^{k}$ the present data combined with previous studies
suggest $k<3$. The $k$ values are constrained by the relatively low
fraction of the soft X-ray background that remains unresolved by deep
surveys (6--26\%). Higher $k$ values will result in an overproduction
of the soft X-ray  background given the fraction already attributed to
AGN, groups and  clusters. The mean X-ray emissivity of spiral galaxies
at $z\approx0.1$ is also estimated and is found to be consistent
within the uncertainties with that of local H\,II galaxy samples. 
Using the mean X-ray emissivity of the spiral galaxy sub-sample we
estimate a global star-formation density of $0.009\pm0.007\rm
M_{\odot}\,yr^{-1}$ at $z\approx0.1$. Although the uncertainty is
large this is lower than previous results based on galaxy samples
selected at different  wavelengths. Nevertheless, given the large
uncertainties involved in converting X-ray luminosity to
star-formation rate the agreement is surprising.  
\end{abstract}


\section{Introduction}\label{sec_intro}

The diffuse X-ray background (XRB), first discovered by
Giacconi et al. (1962), is believed to be dominated by obscured and
unobscured  AGNs and QSOs.  Indeed, deep {\it ROSAT} surveys
to the limit $\approx 10^{-15}\rm \, erg \, s^{-1}\, cm^{-2}$ have 
resolved 60-80 per cent of the soft band (0.5-2\,keV) XRB into
discrete sources (e.g. Boyle et al. 1994; Hasinger et al. 1998;
Lehmann et al. 2001).
Follow-up spectroscopic studies have demonstrated that most of them
are AGNs with a small fraction associated with groups or clusters
(Georgantopoulos et al. 1996; Schmidt et al. 1998;  
Lehmann et al. 2001). Recently,  sensitive {\it Chandra}   surveys
extended the  studies above to fainter flux limits, $\approx
10^{-17}-10^{-16}\rm \, erg \, s^{-1}\, cm^{-2}$, resolving 80-95 per
cent of the soft band XRB (Mushotzky et al. 2000; Giacconi et
al. 2002; Rosati et al. 2002). A key development from these ultra-deep 
surveys is the emergence of X-ray faint sources with properties
similar to those of `normal' galaxies (i.e. systems not obviously
dominated by AGN 
activity).  These sources, although too faint to dominate  the XRB
($<40$ per cent; Giacconi \& Zamorani 1987; Griffiths \& Padovani
1990; Pearson et al. 1997; Alexander et al. 2002; Hornschemeier et
al. 2002a; Persic \& Rephaeli 2003; Ranalli, Comastri \& Setti 2003)
are likely to outnumber AGNs  at faint flux  limits ($<10^{-17}\rm \,
erg\,sec^{-1}\,cm^{-2}$;  Hornschemeier et al. 2002b; Ranalli et
al. 2003). Study of this population to high redshifts provides a
census of the global star-formation rate complementary to that
obtained from other wavelengths. More importantly, the X-rays are
unique in that they are the only tool available to probe the evolution
of low- and high-mass X-ray binaries as well as hot gas. Comparison of
the X-ray properties of local `normal' galaxy samples with high
redshift ones provides a picture of the evolution of the Universe that
is complementary to that inferred from the study of QSOs that dominate
the XRB.         

Study of `normal' galaxies in the local Universe ($<100$\,Mpc) has
been initiated by both the {\it Einstein} and the {\it ROSAT}
satellites. Ellipticals were found to retain large amounts ($\approx
10^{8}-10^{11} M_\odot$) of hot gas ($T\approx10^{7}$\,K) that dominate
their X-ray emission (Fabbiano 1989; Kim, Fabbiano \&
Trinchieri 1992; Fabbiano \& Shapley 2002),  while the spiral galaxy
X-rays were shown to originate from a combination of stellar point
sources and diffuse hot ($T \approx 1-8\times 10^{6}$\,K) gas (Read,
Ponman \& Strickland 1997; Read \&  Ponman 2001). Moreover, a number
of studies have demonstrated  the significance of the environment, the 
star-formation history and the total galaxy mass in understanding the
`normal' galaxy X-ray emission (Read \& Ponman 2001; O'Sullivan,
Forbes \& Ponman 2001;  Helsdon et al. 2001; Fabbiano \& Shapley 2002).  

In the last few years the launch of XMM-{\it Newton} and {\it Chandra},
with their significantly improved sensitivity and resolution,
enabled study of the X-ray properties of `normal' galaxies  at
cosmological distances. Brandt et al. (2001b) studied the mean X-ray
properties of bright spirals at $z\approx0.5$ using a 500\,ks {\it
Chandra} exposure of the HDF-North. They  found an increase in the
mean X-ray  luminosity of their sample compared to local spirals
suggesting evolution. Hornschemeier et al. (2002a) used stacking   
analysis to investigate the mean properties of $z\approx1$ spirals in
the {\it Chandra} Deep Field North. They estimated a mean 0.5-2\,keV
flux of  $\approx5\times 10^{-18}\rm{ erg\,cm^{-2}\,s^{-1}}$ for the
spirals  in their sample and found evidence for an increase in their 
$L_X/L_B$ ratio. However, statistical uncertainties did not allow firm
conclusions to be drawn. Stacking of Lyman break galaxies in the same
field extended our knowledge of the X-ray properties of non-AGN
dominated galaxies to $z\approx3$ (Brandt et al. 2001a; Nandra et
al. 2002). Lyman break systems, believed to be a combination of
starbursts and   low luminosity AGNs, are found to have mean X-ray
luminosities and $L_X/L_B$ ratios comparable to those of the most
luminous starbursts in the local universe (Brandt et al. 2001a). 

Although the studies above provide valuable information on the nature
and the evolution of the X-ray emission in `normal' galaxies, they
focus on either local ($\approx20$\,Mpc) or higher  redshift ($z>0.4$)  
samples. Study of the X-ray properties of `normal' galaxies in the 
intermediate redshift regime is still missing although critical in
interpreting their evolutionary properties by providing a link between
local and  distant samples. In this paper we address this issue using
a shallow (2-10\,ksec) XMM-{\it Newton} X-ray survey covering an area
of $\approx1.5\,\rm deg^{2}$. These observations overlap with the 2dF
Galaxy Redshift Survey (2dFGRS; mean redshift $z\approx0.1$; Colless
et al. 2001) providing a unique opportunity to study the  mean X-ray
properties of `normal' galaxies at  $z\approx0.1$. Compared to
previous studies, the 2dFGRS has the advantage 
of high quality spectra, accurate redshift measurements and reliable
classifications into different galaxy types (E/S0, spirals) based on
the observed spectroscopic properties.  Moreover, sources with optical
spectra revealing powerful AGNs can easily be discarded from the
`normal' galaxy sample.  Our analysis of the mean  X-ray properties of
a well defined sample of `normal' galaxies provides a reference point
at $z\approx0.1$ for comparison with similar studies at higher
redshifts. 

Section \ref{intro} presents the XMM/2dF survey, section \ref{obs}
describes the reduction of the X-ray data while section \ref{sample}
details the selection criteria of the sample used in the present
study. The stacking technique is outlined in  section \ref{stacking},
the results are presented in section \ref{results} and are discussed
in section \ref{discussion}.  Finally section \ref{conclusions}
summarises  our conclusions.  Throughout this paper we adopt
$\rm{H_{o}=65\,km\,s^{-1}\,Mpc^{-1}}$ and $q_o=0.5$. 

\section{The XMM/2dF Survey}\label{intro}
The North Galactic Pole F864 region (RA(J2000)=$13^{\rm h}41^{\rm m}$;
Dec.(J2000)=$00\degr00\arcmin$) and the South Galactic Pole
(SGP; RA(J2000)=$00^{\rm h} 57^{\rm m}$, Dec.(J2000)=$-28\degr
00\arcmin$) were surveyed by XMM-{\it Newton} between May and July 
2002 as part of the Guaranteed Time program. The observations consist
of 9 pointings in the SGP and 7 pointings in F864 region each with an  
exposure time of $\approx2-10$\,ksec. The RA and DEC of the center of
each pointing are listed in Table 1. The EPIC (European Photon Imaging 
Camera; Str\"uder et al. 2001; Turner et al. 2001) cameras were
operated in full frame mode with the thin filter applied.   

Both the F864 and SGP regions overlap with the
2dFGRS\footnote{http://msowww.anu.edu.au/2dFGRS/} (Colless et 
al. 2001). The 2dFGRS is a large-scale spectroscopic survey that fully
exploits the capabilities of the 2dF multi-fiber spectrograph on the
the 4\,m Anglo-Australian Telescope (AAT). This is an on-going
project that aims to obtain high quality spectra and redshifts for
250\,000 galaxies brighter than an extinction corrected limit of
$bj<19.4$\,mag over an area $\approx 2000$\,deg$^2$. Recently, data
for about 100\,000 galaxies have been released (100k data
release). Part of this unprecedented spectroscopic database overlap
with our XMM-{\it Newton} survey and hence optical spectroscopic
information is available for galaxies in the F864 and SGP regions to
the above  magnitude limit.  Moreover, the central regions of both the
F864 and SGP XMM-{\it Newton} survey have been observed by the {\it
ROSAT} satellite (Griffiths et al. 1996; Georgantopoulos et al. 1996)

\section{XMM-Newton Observations}\label{obs}
The {\it XXM-Newton} data have been analysed using the Science
Analysis Software (SAS 5.3). Event files for the PN and the two MOS
detectors have been produced using  the {\sc epchain} and {\sc
emchain} tasks of SAS respectively. The event files were screened for
high particle  background periods by rejecting times with 0.5-10\,keV
count rates higher than 25 and 15\,cts/s for the PN and the two MOS
cameras respectively. These criteria excluded 5 fields (F864--4,
SGP--1, 5, 7  and 9) from the analysis that suffered by significantly
elevated and flaring particle background. The PN and MOS good time
intervals for the remaining pointings are shown in Table 1. The
differences between the PN and MOS exposure times are due to varying
start and end times of individual observations. Only events 
corresponding to patterns  0--4 for the PN and 0--12 for two MOS
cameras have been kept. To increase the signal--to--noise ratio and to
reach fainter fluxes the PN and the MOS event files have been combined
into a single event list using the {\sc merge} task of SAS.

Images in celestial coordinates with pixel size of 4.35\,arcsec have
been extracted in the spectral bands 0.5-2\,keV (soft) and 2-8\,keV
(hard) for the merged event file. Exposure maps accounting for
vignetting, CCD gaps and bad pixels  have been constructed for each
spectral band. 

Source detection is performed on the 0.5-8\,keV image using the {\sc
wavdetect} task of CIAO with a false probability threshold of
$10^{-5}$ (Freeman et al. 2002) corresponding to about $5\sigma$. A
total of 508 X-ray sources have been detected to the limit $f_X(\rm
0.5-8\,keV)\approx10^{-14}\,erg\,s^{-1}\,cm^{-2}$. The 0.5-8\,keV
detection flux limit of individual XMM-{\it Newton} pointings for a
point source on-axis are shown in Table 1. We note that the false
probability threshold of $10^{-5}$ for the total band (0.5-8\,keV)
corresponds to a detection threshold of about $4\sigma$ in the soft
band (0.5-2\,keV). This limit translates to a  typical on-axis source
flux of $f_X(\rm 0.5-2\,keV)\approx4 \times
10^{-15}\,erg\,s^{-1}\,cm^{-2}$ in the soft 0.5-2\,keV band.

A byproduct of the source extraction algorithm is the construction of
background maps for each spectral band. The mean background counts per
pixel (estimated from the background maps) are also listed in Table
1. A detailed analysis of the nature of the X-ray sources detected in
the F864 and SGP areas will be presented in a forthcoming paper
(Georgantopoulos et al. in preparation). To convert counts to flux the
Energy Conversion Factors (ECF) of individual detectors are calculated
assuming a power law spectrum with $\Gamma=1.7$ and Galactic
absorption $N_H=2\times 10^{20} \rm {cm^{-2}}$ appropriate for both
the SGP and F864 fields. The mean ECF for the mosaic of all three
detectors is estimated by weighting the ECFs of individual detectors
by the respective exposure time.  For the encircled energy correction,
accounting for the energy fraction outside the aperture within which
source counts are accumulated, we adopt the calibration performed by
Ghizzardi (2001a, 2001b). These studies use both PN and MOS
observations of point sources to formulate the XMM-{\it Newton} PSF  
for different energies and off-axis angles. In particular, a King
profile is fit to the data with parameters that are a function
of both energy and off-axis angle. 

\begin{table*} 
\footnotesize 
\begin{center} 
\begin{tabular}{ccccc cc} 
\hline 
Field Name & RA          & Dec       & PN exp. time & MOS exp. time &
$f(\rm 0.5-8)$$^{a}$ & background$^{b}$ \\  
           & (J2000)   & (J2000) &     (sec) &  (sec) &  & (cnts) \\
\hline 
F864-1
& $13\mathrm{^h} 41\mathrm{^m} 24.0\mathrm{^s}$ 
& $+00\mathrm{^\circ} 24\mathrm{^\prime} 00\mathrm{^{\prime\prime}}$ 
& 5779  & 9974 & 1.0 & 0.33 \\
F864-4  
& $13\mathrm{^h} 41\mathrm{^m} 24.0\mathrm{^s}$ 
& $+00\mathrm{^\circ} 00\mathrm{^\prime} 00\mathrm{^{\prime\prime}}$ 
& -- & -- & -- & --\\
F864-5
& $13\mathrm{^h} 43\mathrm{^m} 00.0\mathrm{^s}$ 
& $+00\mathrm{^\circ} 00\mathrm{^\prime} 00\mathrm{^{\prime\prime}}$ 
& 1693 & 4447 & 2.8 & 0.40 \\
F864-6
& $13\mathrm{^h} 44\mathrm{^m} 36.0\mathrm{^s}$ 
& $+00\mathrm{^\circ} 00\mathrm{^\prime} 00\mathrm{^{\prime\prime}}$ 
& 2766 & 6493 & 1.6  & 0.29 \\
F864-7
& $13\mathrm{^h}$ $41\mathrm{^m} 24.0\mathrm{^s}$ 
& $-00\mathrm{^\circ} 24\mathrm{^\prime} 00\mathrm{^{\prime\prime}}$ 
& 3459 & 7139 & 1.4 & 0.27 \\
F864-8
& $13\mathrm{^h} 43\mathrm{^m} 24.0\mathrm{^s}$ 
& $-00\mathrm{^\circ} 24\mathrm{^\prime} 00\mathrm{^{\prime\prime}}$ 
& 2109 & 7276 & 2.2 & 0.62 \\
F864-9
& $13\mathrm{^h} 44\mathrm{^m} 36.0\mathrm{^s}$ 
& $-00\mathrm{^\circ} 24\mathrm{^\prime} 00\mathrm{^{\prime\prime}}$ 
& -- & 7725 &  3.6 & 0.46 \\	
SGP-1     
& $00\mathrm{^h} 55\mathrm{^m} 19.0\mathrm{^s}$ 
& $-27\mathrm{^\circ} 36\mathrm{^\prime} 00\mathrm{^{\prime\prime}}$ 
& -- & -- & -- & -- \\
SGP-2  
& $00\mathrm{^h} 57\mathrm{^m} 00.0\mathrm{^s}$ 
& $-27\mathrm{^\circ} 36\mathrm{^\prime} 00\mathrm{^{\prime\prime}}$ 
& 4267 & 8061 & 1.1 & 0.36 \\
SGP-3
& $00\mathrm{^h} 58\mathrm{^m} 51.0\mathrm{^s}$ 
& $-27\mathrm{^\circ} 36\mathrm{^\prime} 00\mathrm{^{\prime\prime}}$ 
& 3181 & 7729 & 1.1 & 0.20 \\
SGP-4
& $00\mathrm{^h} 55\mathrm{^m} 19.0\mathrm{^s}$ 
& $-28\mathrm{^\circ} 00\mathrm{^\prime} 00\mathrm{^{\prime\prime}}$ 
& 4428& 8286 & 1.1 & 0.26 \\
SGP-5
& $00\mathrm{^h} 57\mathrm{^m} 00.0\mathrm{^s}$ 
& $-28\mathrm{^\circ} 00\mathrm{^\prime} 00\mathrm{^{\prime\prime}}$ 
& -- & -- & -- & --\\
SGP-6
& $00\mathrm{^h} 58\mathrm{^m} 51.0\mathrm{^s}$ 
& $-28\mathrm{^\circ} 00\mathrm{^\prime} 00\mathrm{^{\prime\prime}}$ 
& 6390 & 10358 & 1.0 & 0.35 \\
SGP-7
& $00\mathrm{^h} 55\mathrm{^m} 19.0\mathrm{^s}$ 
& $-28\mathrm{^\circ} 24\mathrm{^\prime} 00\mathrm{^{\prime\prime}}$ 
& -- & -- & -- & --\\	
SGP-8
& $00\mathrm{^h} 57\mathrm{^m} 00.0\mathrm{^s}$ 
& $-28\mathrm{^\circ} 24\mathrm{^\prime} 00\mathrm{^{\prime\prime}}$ 
& 7787 & 3456 & 0.9 & 0.32 \\	
SGP-9
& $00\mathrm{^h} 58\mathrm{^m} 51.0\mathrm{^s}$ 
& $-28\mathrm{^\circ} 24\mathrm{^\prime} 00\mathrm{^{\prime\prime}}$ 
& -- & -- & --\\	
\hline 
\multicolumn{7}{l}{$^a$Limiting 0.5-8\,keV flux for detection
($5\sigma$) in units
of $\rm 10^{-14}\,erg\,s^{-1}\,cm^{-2}$} \\
\multicolumn{7}{l}{$^b$Mean 0.5-8\,keV background counts per pixel
from the background maps produced by the {\sc wavdetect} task.} \\

\end{tabular} 
\end{center} 
\caption{Observing Log  of XMM-{\it Newton} F864 and SGP survey}   
\normalsize  
\end{table*}

\section{Galaxy sample}\label{sample}
The `normal' galaxy sample used in the present study is compiled from
the 100k data release of the 2dF Galaxy Redshift 
Survey. A full description of the survey design, spectroscopic
observations, data reduction and redshift measurements  of the 2dFGRS
dataset can be found at Colless et al. (2001). In brief, redshift
estimation is performed using two independent methods: the first
cross-correlates the spectra with a library of absorption-line
galaxy and stellar templates after clipping emission lines, while the 
second finds and then fits emission lines. The automated redshift
estimates are also confirmed by visual inspection. Occasionally, a
manual redshift is determined by fitting spectral features missed by
the automated method (Colless et al. 2001).  A quality flag, $Q$,
measuring the reliability of the estimated redshifts  has been
assigned. Reliable redshifts have $Q\ge3$, while $Q=2$ corresponds to
a probable redshift and $Q=1$ indicates no redshift measurement.

Spectral classification of the 2dFGRS galaxies has been performed
by applying Principal Component Analysis (PCA) techniques on the
observed spectra (Folkes et al. 1999; Madgwick et al. 2002). Spectral
types are  determined on the basis of the 
$\eta$ parameter: 
\begin{equation}
 \eta=0.5\times pc_1-pc_2,
\end{equation} 
where $pc_1$, $pc_2$ are the projections of the first two eigenspectra
derived from PCA. Comparison of the $\eta$ parameter of the 2dFGRS
galaxies with that derived for the observed spectra of the Kennicutt
galaxy atlas (Kennicutt 1992) shows that there is a continuous
relation between $\eta$ and morphological types (Madgwick et al. 2002,
their Figure 4). We adopt the classification scheme of Madgwick et
al. (2002) and group galaxies  into four spectral types: E/S0, Sa, Sb
and Scd. The adopted range of the $\eta$ parameter for the above
spectral types are presented in Table \ref{tab_eta}. ``Unclassified''
galaxies with no spectral classification are systems with low
signal--to--noise spectra that did not allow spectral type
determination.  

\begin{figure} 
\centerline{\psfig{figure=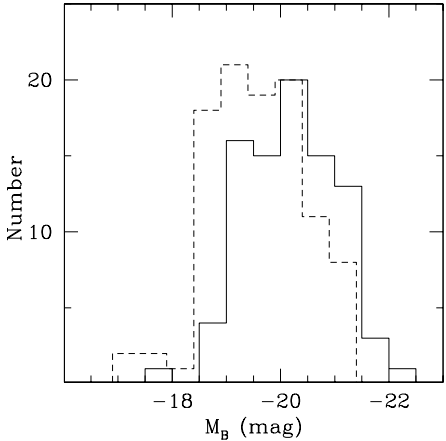,width=2.5in,height=2.5in,angle=0}} 
\caption
 {$B$-band absolute magnitude distribution of the spiral (dashed line)
 and the elliptical (continuous line) 2dFGRS sub-samples. For clarity
 the spiral galaxy histogram is shifted to the left by 0.1
 magnitudes. 
 }\label{fig_mag_dist}     
\end{figure}

The present sample comprises a total of 235 2dFGRS galaxies with $Q>2$
that lie within 13\,arcmin off the center of the EPIC-PN camera to  
avoid  the low signal--to--noise areas at the edge of the XMM-{\it
Newton} field of view. Visual inspection of the mosaiced images
removes three galaxies close to CCD gaps. Moreover, two 2dFGRS sources
(TGN\,335Z095 and TGN\,336Z241) lie close to an extended  X-ray
source, probably a cluster. These  galaxies are not individually
detected at X-ray wavelengths and are excluded from the stacking
analysis 
described in section \ref{stacking} to avoid contamination of the
stacked signal by  emission not directly associated to the galaxies. A
further 4 sources with stellar optical spectra are also excluded from
the analysis reducing  the final sample of  2dFGRS galaxies to
226. The distribution of the present sample of 2dFGRS galaxies to
spectral types as well as the corresponding mean redshifts are shown
in Table \ref{tab_eta}. We also note that the present sample is
$\approx70$ per cent complete. The incompleteness is due to $Q<2$
galaxies as well instrumental limitations (e.g. fiber collisions).  
The optical $B$-band absolute magnitude distributions of ellipticala
and spirals in the sample are
shown in Figure \ref{fig_mag_dist}.

\begin{table} 
\footnotesize 
\begin{center} 
\begin{tabular}{ccccc} 
\hline 
Spectral & $\eta$  & Number & $<z>$ & $\sigma_{z}$ \\
type     &         & of galaxies   &  & \\ 
\hline
 all   &  --                  &  226   & 0.110 & 0.003 \\
 E/S0  & $\eta<-1.4$          &   88   & 0.105 & 0.004 \\
 Sa    & $-1.4\le \eta < 1.1$ &   59   & 0.104 & 0.005 \\
 Sb    & $ 1.1\le \eta < 3.5$ &   28   & 0.092 & 0.005 \\ 
 Scd   & $\eta \ge 3.5$       &   19   & 0.085 & 0.007 \\
Unclassified & --             &   32   & 0.139 & 0.013 \\
\hline 
\end{tabular} 
\end{center} 
\caption{Spectral classification scheme for the 2dFGRS galaxies}\label{tab_eta}  
\normalsize  
\end{table}

\section{Stacking procedure}\label{stacking}
Stacking methods have been extensively used in X-ray astronomy to
study  the mean properties (e.g. flux, luminosity, hardness ratios) of
well defined samples of sources that are otherwise too  faint at
X-rays to be individually detected (e.g. Nandra et al. 2002).   

In practice the X-ray counts (source--plus--background) at the
position of each galaxy in the sample are added excluding 
X-ray detected galaxies. The expected background contribution is
estimated by summing the counts from regions around each
source. Assuming Poisson statistics for the counts a significance
level is estimated  for the summed signal. 
 
To determine the size of the region within which the
source--plus--background counts are extracted we adopt the empirical 
method described by Nandra et al. (2002). Tests are performed in
which the radius of the circular aperture within which the 
source--plus--background counts are summed varies (from one trial to
the next) between 8--18\,arcsec. We find that a radius of 12\,arcsec
maximises the significance of the stacked signal and hence
this optimal extraction radius is adopted for the analysis that
follows. This extraction radius is about 3 times the on-axis HWHM of
the XMM-{\it Newton} PSF (Hasinger et al. 2001). We also note that the
visible disk of a Milky Way type galaxy at z=0.1 subtends an angular
radius of $\approx8$\,arcsec.       

To assess the significance of the stacked signal we estimate the
background by averaging the counts in 60 12\,arcsec apertures 
randomly  positioned within annular regions centered on the positions
of the sample galaxies with inner and outer radii of 20 and
100\,arcsec respectively. To avoid contamination from X-ray detections
in the background estimation, regions close to X-ray sources
($8-40$\,arcsec) are excluded. 
The smooth background map produced  by the {\sc wavdetect} task of
CIAO was also used to estimate the background. Although no significant
differences are found, the background estimated in the latter case is
systematically lower and hence, the detection significance of the
stacked signal is somewhat higher. We adopt a conservative approach and
estimate the background from the science image rather than the smooth
background map. However, it should be noted that use of the
background map does not alter any of the conclusions.

In practice, 2dFGRS galaxies associated ($<8$\,arcsec)
with X-ray sources detected in the total (0.5-8\,keV) band above the
false probability threshold of $10^{-5}$ are  excluded from
the stacking analysis. As already discussed in the previous section
this detection limit corresponds to a soft band (0.5-2\,keV) detection
threshold of about $4\sigma$ which translates to a typical on-axis
point source flux of $f\rm (0.5-2\,keV)\approx4\times10^{-15}\rm\,
erg\,  s^{-1}\, cm^{-2}$. Such a low  detection threshold is essential
to study the mean properties of X-ray weak sources, the signal of
which would otherwise be diluted by brighter ones. 


X-ray detected 2dFGRS galaxies may differ in their  nature
from X-ray faint galaxies. It is clear that summing counts of
different classes of objects is meaningless. We find a total of 7
2dFGRS galaxies associated ($<8$\,arcsec) with X-ray detections 
above the total band (0.5-8\,keV) false probability threshold of
$10^{-5}$. Two of them exhibit clear  evidence for AGN activity such
as  broad emission  
line optical spectra, high X-ray--to--optical luminosity ratios
($L_X/L_B>-1$) and X-ray luminosities $L_X\approx10^{42}\,\rm
erg\,s^{-1}$. Another three 2dFGRS galaxies have narrow emission line
optical spectra, $L_X/L_B\approx-1$, relatively hard X-ray 
spectral properties and soft band (0.5-2\,keV) X-ray luminosities 
in the range $0.3-5 \times 10^{42} \,\rm erg\,s^{-1}$. The
evidence above suggests obscured or weak AGN activity. Finally two
2dFGRS galaxies have optical spectra exhibiting both absorption and
narrow emission lines, $L_X/L_B\approx-2$, soft X-ray spectra  and 
$L_X\approx2 \times 10^{41}\,\rm erg\,s^{-1}$, suggesting either
`normal' galaxies or LLAGNs. Detailed analysis of these galaxies will
be presented in a forthcoming paper (Georgakakis et al. in
preparation). 

Finally, counts associated with the wings of the PSF of
bright X-ray sources might erroneously increase the stacking signal
significance. Therefore, to avoid contamination of the stacking signal
from nearby X-ray sources we have excluded from the analysis 2dFGRS
galaxies that lie close ($<40$\,arcsec) to X-ray detections
(total of 21). Therefore, a total of 28 2dFRGS sources either lie
close to (21; $8-40$\,arcsec) or are associated with (total of 7;
$<8$\,arcsec) X-ray detections. This reduces the number of galaxies
used in the stacking analysis to 198.   

\section{Results}\label{results}

\begin{table*} 
\footnotesize 
\begin{center} 
\begin{tabular}{ccc ccc ccc ccc cc} 
\hline 
spectral 
& source
&  median
& \multicolumn{2}{c}{S+B$^{a}$} 
& \multicolumn{2}{c}{B$^{b}$}  
& \multicolumn{2}{c}{SNR$^{c}$} 
& \multicolumn{2}{c}{$f_X$$^{d}$} 
& \multicolumn{2}{c}{$L_X$$^{e}$} 
& $L_B$$^{f}$
\\

type 
& number 
& $z$
& SB & HR  
& SB & HR   
& SB & HR 
& SB & HR
& SB & HR
& \\


\hline

All 
& 198
& 0.090
& 824 & 807
& 676.5 & 793.2
& 5.7 & 0.5
&$7.4\pm2.3$ & $<14.0$
& $1.6\pm 0.5$  & $<3.0$
& $6.6$
\\

E/S0 
& 83 
& 0.089
& 371 & 280
& 299.3 & 306.7
& 4.1 & --
&$14.5\pm 4.7$ & $<21.4$ 
&$3.1\pm 1.0$  & $<4.5$
& $8.5$
\\

Sa-Scd  
& 87 
& 0.088
& 374 & 424
& 351 & 393.8
& 3.3 & 1.6
&$2.5 \pm  2.2$ & $<27.8$ 
&$0.5\pm 0.4$  & $<2.2$
& $5.3$
\\
\hline
\multicolumn{14}{l}{$^a$Source+Background counts in the soft (SB) and
hard (HB) bands} \\
\multicolumn{14}{l}{$^b$Background counts in the soft (SB) and
hard (HB) bands} \\
\multicolumn{14}{l}{$^c$Significance of detection in background standard
deviations in the soft (SB) and hard the (HB) bands} \\
\multicolumn{14}{l}{$^d$X-ray flux in units of $\rm
10^{-16}\,erg\,s^{-1}\,cm^{-2}$ in the soft (SB) and hard
the (HB) bands} \\
\multicolumn{14}{l}{$^e$X-ray luminosity in units of $\rm
10^{40}\,erg\,s^{-1}$ in the soft (SB) and hard
the (HB) bands} \\
\multicolumn{14}{l}{$^f$B-band luminosity in units of $\rm
10^{42}\,erg\,s^{-1}$} \\

\end{tabular} 
\end{center} 
\caption{
Stacking analysis results in the soft (0.5-2\,keV) and hard
(2-8\,keV) bands for different samples of 2dFGRS
galaxies}\label{stacking_tbl} 
\normalsize  
\end{table*} 

The stacking results for the 2dFGRS galaxies are summarised in  Table
\ref{stacking_tbl}. The X-ray flux in this table is estimated from the
raw net counts after correcting individual galaxies for (i) the effect 
of vignetting estimated from the corresponding exposure map 
and (ii) the energy fraction outside the adopted aperture of
12\,arcsec using the encircled energy corrections recently derived by
Ghizzardi  (2001a, 2001b) for both the PN
and the MOS detectors (see section \ref{obs}). The counts--to--flux
conversion factor has been derived assuming a power-law with spectral
index $\Gamma=1.7$ and Galactic absorption (see section \ref{obs}).
The X-ray luminosities in Table \ref{stacking_tbl} are estimated
using the X-ray flux derived from the stacking analysis and the mean
redshift of each subsample also listed in this table. 

The soft-band yields a statistically significant signal for both the
whole sample and for the E/S0 and spiral galaxy sub-samples. On the
contrary stacking of the hard band  counts did not result in a
significant signal for either of the samples. For the 2-8\,keV band
$3\sigma$ upper limits are estimated assuming Poisson statistics for
the background stacked counts. Unfortunately, the estimated upper 
limits are not tight enough to strongly constrain the X-ray spectral
properties of the spiral and elliptical galaxy subsamples.   

To further assess the confidence level of the stacked signal in both
the soft and the hard bands  we perform extensive simulations: mock
catalogs are constructed by randomising the positions of the
2dFGRS samples in Table \ref{stacking_tbl} avoiding areas close to
X-ray sources ($<40$\,arcsec). Each of the mock catalogs
has the same number of  sources as the  2dFGRS samples in Table
\ref{stacking_tbl}. We then apply the stacking analysis technique
outlined in section \ref{stacking} to the mock catalogs and estimate
the significance of the stacked signal in the same way as in the real
catalog. This is then repeated 10\,000 times to get the distribution
of the detection significances (as defined in section \ref{stacking})
for the mock catalogs. This provides an estimate of the probability
of getting a statistically significant stacked signal by chance. The
detection confidence levels estimated from the simulations are in
excellent agreement with those shown in Table \ref{stacking_tbl} based
on Poisson statistics



We also investigate the possibility that the observed stacked signal
is primarily due to a few bright X-ray sources below the detection
threshold. This is important especially if some of these sources are
AGNs and hence, not representative of the `normal' galaxy population
studied in this paper. Firstly, we find no evidence of broad emission
lines in the optical spectra of the galaxies used in the stacking
analysis. Therefore, the obtained stacked signal does not include any
contribution from powerful AGNs/QSOs, although we cannot exclude the
possibility of weak and/or heavily obscured  AGNs present in the 
sample. We further explore this  issue by assigning detection
significances to individual 2dFGRS and then excluding sources
above a given threshold. The detection significance for each source is
defined as the ratio of net source counts to the square root of the
background.  We note that this definition of detection  significance is
different to that used  by the {\sc wavdetect} task. 

For the whole sample we need to remove galaxies above $2.7\sigma$ 
(total of 11) to obtain a low confidence level signal
($2.6\sigma$). The 11 sources that appear to dominate the stacking
signal have mean $f_X/f_B\approx-2$ and $L_X(\rm
0.5-2\,keV)\approx10^{41}\rm erg\,s^{-1}$ suggesting  
that the  contribution from AGNs is small. Moreover, the optical 
spectra of the systems that appear to dominate the stacking signal
do not show any evidence for AGN activity (e.g. broad or high
excitation lines). The corresponding cutoffs for a low confidence
stacked signal in the spiral and elliptical  subsamples is
$3\sigma$. Deeper  X-ray observations are required to further
address this issue.    

Finally, we note that for the stacked signal we estimate an 
X-ray--to--optical flux ratio  $\log f_X / f_B \approx-2.5$ typical of
`normal' galaxies and $L_X\approx10^{40}\rm \,erg\,sec^{-1}$ similar
to that of Milky Way type galaxies. This also suggests that any 
contribution from AGNs in the stacking analysis is small.
However, we note that the present data cannot be used to identify and
exclude from the stacking analysis weak or heavily obscured AGNs that
do not manifest their presence in the optical galaxy spectra. For
example, Comastri et al. (2002) have identified  a class of X-ray
bright optically inactive galaxies with optical spectra
typical of `normal' early type systems and X-ray properties indicative
of obscured AGN activity. Our sample is likely to be
contaminated by such systems.  We note that because of their early
type optical spectra sources like those found by Comastri et
al. (2002) are likely to primarily affect the elliptical rather than
the spiral galaxy subsample.

\begin{figure} 
\centerline{\psfig{figure=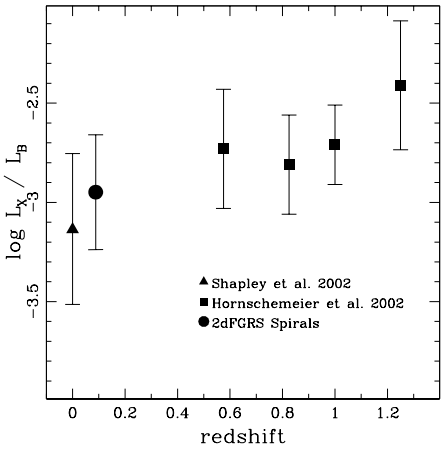,width=2.5in,height=2.5in,angle=0}} 
\caption
 {$\log L_X / L_B $ against redshift of spiral galaxies. The
 stacking analysis results for the 2dFGRS spirals are shown with the
 filled circle. The triangle is the median $\log L_X / L_B $ of local
 spirals (Shapley et al. 2001) selected to span an $L_B$ range similar
 to that of the  present sample. For this point the errorbars
 correspond to the 25th and 75th percentile of the $\log L_X / L_B $ 
 distribution of the Shapley et al. (2001) local spiral galaxy
 sample. The squares are   the stacking analysis results  for distant
 late type galaxies in the {\it Chandra} Deep Field North (squares;
 Hornschemeier  2002a). 
 }\label{fig_lxlbz}     
\end{figure}

\section{Discussion}\label{discussion}
\subsection{Optical--to--X-ray flux ratio}
Shapley, Fabbiano \& Eskridge (2001) studied the X-ray
properties of a large sample of local spirals using {\it Einstein}
archival data. They performed multivariate statistical analysis and
report a non-linear correlation between optical and X-ray luminosities 
$L_X\approx L_B^{1.5}$ independent of spectral type. This contradicts
previous studies, based on smaller samples, that found an almost
linear $L_X - L_B$ relation, interpreted as evidence for a link
between X-rays and the evolved stellar population. To explain the 
non-linearity in  $L_X-L_B$ Fabbiano \& Shapley (2002) suggest two
different processes affecting the X-ray emission of early and late
type spirals: the presence of hot ISM similar to that of E/S0 in early
types and dust obscuration in star-forming  regions in late types.   

Brandt et al. (2001b) performed stacking  analysis on a sample of 
17 $z\approx0.5$ `normal' galaxies (both early and late type) using a
500\,ks {\it Chandra} exposure of the HDF-N region. They estimate a mean
0.5-2\,keV luminosity for the spiral galaxies in their sample of 
$L_X=8.5\times10^{39}\rm{erg\,s^{-1}}$ consistent with the  
calculations in the present paper. This $L_X$ is
elevated compared to local spirals but this is attributed to the
increased mean $L_B$ in their sample rather than evolution. 
Similarly, Hornschemeier et al. (2002a) used stacking analysis
techniques to study the X-ray properties of the spiral and irregular
type systems in the {\it Chandra} Deep Field North at redshifts
$z\approx0.4-1.5$. They argue that the mean $L_X/L_B$ in their sample
is elevated compared to that of Shapley et al. (2001)
(selected to span the same $L_B$ range) suggesting evolution. However, the
uncertainties in their $L_X/L_B$ estimates do not allow firm
conclusions to be drawn. 

The present sample with well defined selection criteria is
intermediate in redshift ($z\approx0.1$) to that of Shapley et al.
(2001; $z\approx0.004$) and Hornschemeier et al. (2002a;
$z>0.4$), providing a low-$z$ reference point for comparison with
higher-$z$ studies. Figure \ref{fig_lxlbz}  plots the $\log L_X/L_B$ 
against redshift comparing the results from the studies above. For the
Shapley et al. (2001) spirals the  median $L_X/L_B$ is estimated using
galaxies over the same range of $L_B$ as the present spiral galaxy
subsample. Within the uncertainties there is fair agreement in the
$L_X/L_B$ of local and $z\approx0.1$ 2dFGRS spirals.

Moreover, it is clear from Figure \ref{fig_lxlbz} that the 
$L_X/L_B$ of $z\approx0.1$ spirals in the present study are
also consistent, within the errors, with those of Hornschemeier et
al. (2002a) at $z=0.4-1$. Although the mean $L_X$ of their spirals
increases with redshift from $z\approx0.4$ to $z\approx1$ this is
compensated by a similar increase in the mean $L_B$. At $z>1$
Hornschemeier et al. (2002a) find that both  $L_X/L_B$ and $L_X$ are
elevated compared to our estimates by a factor of 4 suggesting
evolution. Although the large errors hamper a secure interpretation,
it should be noted that this increase, if real, {\it cannot} be 
attributed to differences in $L_B$ between our sample and that of
Hornschemeier et al. (2002a) at $z>1$. Therefore,  our analysis
combined with the results of Hornschemeier et al. (2002a) suggest  
little or no evolution of the spiral galaxy $L_X$ relative to $L_B$ at
least to $z\approx1$. 

\subsection{X-ray emissivity and star-formation rate}
The emissivity of galaxy samples selected at various wavelengths (UV,
optical, radio) over a range of redshifts is one of the most powerful
and widely used evolutionary tests (e.g. Lilly et al. 1996; Connolly et
al. 1997).  The X-ray emissivity, $j_X$, of different galaxy types
(including non-AGN dominated systems) has been estimated locally by 
Georgantopoulos, Basilakos \& Plionis (1999). However, the evolution
of $j_X$ at higher redshifts is still poorly constrained mainly
because `normal' galaxies are faint at X-ray wavelengths and their
detection, especially at high redshift, is difficult even with the
improved sensitivities of the XMM-{\it Newton} and the {\it Chandra}
observatories. 

The 2dFGRS sample at $z\approx0.1$ can provide, for the first time, a 
{\it non}-local estimate of the X-ray emissivity of
`normal' galaxies. For the $j_X$ determination, we first  need to
estimate the effective  cosmological volume probed by the 2dFGRS
survey corrected for optical magnitude incompleteness
(e.g. $b_j<19.4$\,mag) using the standard $1/V_{max}$ formalism
(e.g. Lilly et al. 1996). Corrections have also been applied taking
into account incompleteness of the 2dFGRS sample due to both poor
S/N spectra ($Q<3$, unclassified objects) at faint magnitudes and
instrumental limitations (e.g. fiber collisions) that did
not allow a fraction of the target galaxies to be observed. We further
assume that all 2dFGRS galaxies of a given spectral type have the same
mean $L_X/L_B$ ratio estimated in section \ref{results}. 

For 2dFGRS spirals we estimate a soft-band 0.5-2\,keV emissivity of 
$(0.4\pm0.3)\times 10^{38}$ in fair agreement with that of local
H\,II galaxies, $(0.50\pm0.06)\times 10^{38} {\rm
erg\,sec^{-1}\,Mpc^{-3}}$, estimated by Georgantopoulos et al. (1999;
scaled to the 0.5-2\,keV band). For the elliptical galaxy sub-sample
we find $j_X=(1.2\pm0.4)\times 10^{38} {\rm
erg\,sec^{-1}\,Mpc^{-3}}$. Although the uncertainties are large this 
is higher than the $j_X$ of local passive galaxies, 
$(0.53\pm0.06)\times 10^{38} {\rm erg\,sec^{-1}\,Mpc^{-3}}$,
estimated by Georgantopoulos   et al. (1999).  These authors
convolved the local optical luminosity function of the Ho, Filippenko
\& Sargent (1995) sample with the corresponding $L_X-L_B$ relation
based on {\it Einstein} data. The Ho et al. (1995) sample has the
advantage of high S/N nuclear optical spectra and hence, reliable  
spectral classifications (Ho, Filippenko \& Sargent 1997) necessary to
study the properties of non-AGN galaxies. 

Assuming the same mean 0.5-2\,keV X-ray flux (estimated from stacking;
see Table 1) for all
2dFGRS spiral galaxies and an incompleteness of $\approx70$ per cent
(see section 3) we estimate their XRB contribution to be $\approx0.4$
per cent. 
This calculation assumes that the spectrum of the diffuse X-ray
background follows the form $9\times E^{-0.4} \rm
\,keV\,cm^{-2}\,s^{-1}\,sr^{-1}\,keV^{-1}$ (Gendreau et 
al. 1995). This small XRB fraction produced by `normal' spirals at
$z\approx0.1$ is broadly consistent with studies of `normal' galaxies
at higher redshifts using deep X-ray surveys. Hornschemeier et
al. (2002a) argue that they have identified $\approx1$ per cent of the
soft XRB as arising from spirals not individually detected in deep {\it
Chandra} surveys. Alexander et al. (2002)  studied X-ray detected
starbursts in the redshift range 0.4--1.3 and argued that they produce
$\approx2$ per cent of the 0.5--8\,keV XRB.   

In addition to the above observational constraints, a number of studies
attempt to predict the total XRB fraction produced by an evolving
population of star-forming  galaxies.  Ranalli et al. (2003) used a
nearby well 
defined sample of star-forming galaxies with available {\it ASCA} and
{\it BeppoSAX} data to derive an almost linear relation between  X-ray
luminosity and radio power. Using this correlation and the radio
source counts at sub-mJy flux densities (believed to be dominated by
starbursts) they predict the contribution of these systems to the
XRB. Integrating their predicted soft band (0.5-2\,keV) X-ray source
counts we estimate a 0.5-2\,keV XRB fraction produced by starbursts
in the range $\approx 10-15$ per cent depending on the faint flux
limit of the integration. Persic \& Rephaeli (2003) carefully modeled
the X-ray spectral energy distribution of nearby star-forming galaxies 
to estimate a 2-10\,keV XRB contribution of $5-11$ per cent 
depending on the assumed  evolution. Although they do not estimate the
soft band XRB contribution the fractions above will increase at softer
energies. Griffiths \& Padovani (1990) converted the $\rm 60-\mu m$
luminosity function of IRAS star-forming galaxies to X-ray wavelengths
and argued that star-forming galaxies with mild evolution could produce
20-30 per cent of the XRB. Similarly, Pearson et al. (1997) used IRAS
star-forming galaxies to fit the X-ray counts at faint flux limits and
estimated that such systems could contribute as much as 30-50 per cent
of the XRB.  The studies above clearly demonstrate that there is
significant uncertainty in the  XRB fraction produced by star-forming
galaxies critically depending on the adopted evolution.  

The present study of spiral galaxies to $z\approx0.1$ cannot on its
own be used to either constrain the evolution of these systems with
$z$ or estimate their {\it total} XRB contribution extrapolated to
high $z$. However, comparison of our results at $z\approx0.1$ with
those obtained at higher redshifts (e.g. Hornschemeier et al. 2002)
indicate X-ray evolution for these systems. Indeed, the almost
constant $L_X/L_B$ versus $z$ relation in Figure \ref{fig_lxlbz} can
be interpreted as evidence  that $L_X$ evolves in at least the same
rate as $L_B$.  
It is an interesting exercise to attempt to constrain the X-ray
evolution of star-forming galaxies using the X-ray emissivity of
2dFGRS spirals at $z\approx0.1$ and assuming that the bulk of the soft 
XRB that remains unaccounted for in deep X-ray surveys is produced by
star-forming processes. Deep ROSAT surveys (Hasinger at al. 1998) to
the limit $\approx10^{-15}\rm erg\,s^{-1}\,cm^{-2}$ have resolved
68-81 per cent of the soft ($<2$\,keV) XRB into discrete sources. Most
of them are AGNs (78 per cent; Schmidt et al. 1998) with a small
fraction being galaxy groups or clusters (6 per cent; Lehmann  et
al. 2001). More recently, deep {\it Chandra} surveys (e.g. Mushotzky
et al. 2000; Brandt et al. 2001c; Giacconi et  al. 2002) extended the
study of the X-ray source population to the limits
$\approx10^{-17}-10^{-16}\rm  erg\,s^{-1}\,cm^{-2}$. 
For example, Mushotzky et al. (2000) resolved 6-13 per cent of  the
soft band XRB to sources also  believed to be AGNs with fluxes in the
range $\approx10^{-16}-10^{-15}\rm  erg\,s^{-1}\,cm^{-2}$.   
Only a small fraction of the XRB (1-2 per cent) is believed   
to be produced by  star-forming  galaxies at the flux limits of the
ultra-deep {\it Chandra} surveys (Hornschemeir et al. 2002a; Alexander
et al. 2002). The evidence above suggests that 74-96 per cent of the
soft band  XRB has been resolved and is produced by AGNs, groups or
clusters allowing a maximum of 6-26 per cent to star-forming
galaxies. We employ  the spiral galaxy $j_X$ at $z\approx0.1$ to
constrain the maximum allowed evolution for these systems.  Under the
assumption that the  X-ray emissivity follows the form $(1+z)^{k}$ to
a maximum formation redshift $z_{\rm max}=4$, we find that the index
$k$ should be $<1.5$ and $<3.2$ for the  XRB fraction produced by
these systems to be  $<6$ and $<26$ per cent respectively. We note
that surveys at optical wavelengths  estimate $k\approx3$ (e.g. Lilly
et al. 1996).  

We stress that the XRB fraction due to star-forming
galaxies quoted above (and hence, the derived evolution parameter $k$)
is an upper limit. Although, a number of studies suggest that as much
as 20-50 per cent of the XRB could be produced by starbursts
(e.g. Griffiths \& Padovani 1990;  Pearson et al. 1997) more recent
estimates suggest modest soft band fractions  (e.g. Persic \& Rephaeli
2003; Ranalli et al. 2003). Deep wide field surveys selecting
star-forming galaxies at higher redshifts using homogeneous and well
defined criteria are required to constrain the X-ray evolution of
these systems and hence, their XRB contribution. This paper provides 
such a sample at low redshifts for comparison with future studies.


Finally, the X-ray emission of spirals not dominated by AGN is
associated to the star-formation activity of the parent galaxy
(Fabbiano \& Shapley 2002; Read \& Ponman 2001). Consequently,  the
X-ray emission of these 
systems can be used to probe both the individual and the global
star-formation rate (SFR). However, the quantitative link 
between X-ray luminosity and  SFR is rather tenuous, depending on
poorly constrained parameters such as the Initial Mass Function and
the nature of the stellar sources responsible for the X-ray
emission (e.g. Ghosh \& White 2001). More theoretical and
observational work is clearly needed in this field. Here, we adopt the
empirical 0.5-2\,keV X-ray luminosity to  SFR conversion derived by
Ranalli et al. (2003). For the X-ray luminosity of the spirals in the
present sample we find a mean SFR of $\approx1.3\,{\rm
M_{\odot}\,yr^{-1}}$ while the star-formation density of the present
sample at $z\approx0.1$ is $\approx0.009\,{\rm
M_{\odot}\,yr^{-1}\,Mpc^{-3}}$. This is
lower than previous estimates using galaxy samples selected at 
different wavelengths. However, given the large uncertainties involved
in converting X-ray luminosity to  star-formation rate the 
agreement is surprising. This  suggests that deep X-ray surveys
detecting an  increasing number of `normal' galaxies can be used as
probes of the SFR of individual  systems.   

\section{Conclusions}\label{conclusions}
In this paper we study the mean X-ray properties of `normal' galaxies
using stacking analysis techniques. We use a shallow (2-10\,ksec) XMM-{\it
Newton} survey covering an area  of $\approx 1.5\, \rm deg^{2}$ and
overlapping with the 2dF Galaxy Redshift Survey. The 2dFGRS sample has
the advantage of high quality spectra and hence, accurate redshifts and
reliable classifications into early-type and late-type galaxies. 

In the stacking analysis we use a sample of $\approx200$  2dFGRS
galaxies with a mean redshift of $z\approx0.1$. In the soft 0.5-2\,keV
band we detect a statistically significant signal for both the whole
sample ($\approx6\,\sigma$) and the early ($\approx4\,\sigma$) and late
($\approx3\,\sigma$) type galaxy  sub-samples. On the contrary, no
signal is detected for any of the above samples in the hard 2-8\,keV
band. The main conclusions from this study  are summarised below:
\begin{itemize}

\item We find that the mean $L_X/L_B$ ratio of the present  sample at
$z\approx0.1$ is consistent within the errors with both local
($<100$\,Mpc) and distant ($z\approx1$) samples after accounting for
differences in the mean $L_B$ of the samples. This suggests
little or no evolution of the X-ray emission mechanisms relative to
the optical at least to $z\approx1$. 

\item The mean X-ray  emissivity, $j_X$, of `normal' galaxies at
$z\approx0.1$ is found to be consistent  with that of local H\,II 
samples. We also find that the present sample of 2dFGRS spirals
contributes about 0.4 per cent to the soft band XRB. This low fraction
is in broad agreement with previous studies estimating the XRB
produced by star-forming spirals at moderate redshifts. 

\item 
For star-forming galaxies assuming an X-ray luminosity evolution
of the form  $(1+z)^{k}$ to a maximum redshift $z_{max}=4$ we find
$k<3$. The $k$ values are constrained by the relatively low
fraction of the soft X-ray background that remains unresolved by deep
surveys (6--26\%). Higher $k$ values will result in an overproduction
of the soft X-ray  background given the fraction already attributed to
AGN, groups and  clusters. 

\item Using the mean X-ray properties of the spiral galaxy
sub-sample and assuming an empirical X-ray luminosity to SFR
conversion factor we estimate a global SFR density of
$0.009\pm0.007\rm \, M_{\odot}\,yr^{-1}$ at 
$z\approx0.1$. Although the uncertainties are large this is lower than
previous results  based on galaxy samples
selected at different wavelengths. 
\end{itemize}

\section{Acknowledgments}
 The authors wish to thank the anonymous referee for constructive
 comments. 
 We acknowledge use of the 100k data release of the 2dF Galaxy  
 Redshift Survey. This work is jointly funded by the European Union
 and the Greek Government  in the framework of the programme 'Promotion
 of Excellence in Technological Development and Research', project
 'X-ray Astrophysics with ESA's mission XMM'.

\end{document}